\documentclass[aps,pre,10pt,onecolumn,superscriptaddress,balancelastpage,showpacs,notitlepage]{revtex4-2}
\usepackage[english]{babel}
\usepackage[colorinlistoftodos, color=green!40, prependcaption]{todonotes}

\usepackage{centernot}
\usepackage{graphicx}
\usepackage{amsmath}
\usepackage{times}
\usepackage{amssymb}
\usepackage{mathrsfs}
\usepackage{color}
\usepackage{url}
\usepackage{version}
\usepackage{siunitx}
\usepackage[pdftex,colorlinks=true,pdfstartview=FitV,linkcolor= linkcolor,citecolor= linkcolor,urlcolor= linkcolor,hyperindex=true,hyperfigures=false]{hyperref}
\usepackage{balance}
\usepackage{lastpage}
\usepackage{parskip}
\usepackage{subcaption}
\usepackage[paper=a4paper,margin=1in]{geometry}

\definecolor{linkcolor}{rgb}{0,0,0.6}

\newcommand{\lmax}{\ell_{\mathrm{max}}}

\begin{document}

\title{Thermal fluctuations and osmotic stability of lipid vesicles}

\author{H\aa kan Wennerstr\"om}
\email{hakan.wennerstrom@fkem1.lu.se}
\affiliation{Division of Physical Chemistry, Lund University, P.O. Box 124, S-221 00 Lund, Sweden}

\author{Emma Sparr}
\affiliation{Division of Physical Chemistry, Lund University, P.O. Box 124, S-221 00 Lund, Sweden}

\author{Joakim Stenhammar}
\email{joakim.stenhammar@fkem1.lu.se}
\affiliation{Division of Physical Chemistry, Lund University, P.O. Box 124, S-221 00 Lund, Sweden}


\begin{abstract}
\noindent Biological membranes constantly change their shape in response to external stimuli, and understanding the remodeling and stability of vesicles in heterogeneous environments is therefore of fundamental importance for a range of cellular processes. One crucial question is how vesicles respond to external osmotic stresses, imposed by differences in solute concentrations between the vesicle interior and exterior. Previous analyses of the membrane bending energy have predicted that micron-sized giant unilamellar vesicles (GUVs) should become globally deformed already for nanomolar concentration differences, in contrast to experimental findings that find deformations at much higher osmotic stresses. In this article, we analyze the mechanical stability of a spherical vesicle exposed to an external osmotic pressure in a statistical-mechanical model, including the effect of thermally excited membrane bending modes. We find that the inclusion of thermal fluctuations of the vesicle shape changes renders the vesicle deformation continuous, in contrast to the abrupt transition in the athermal picture. Crucially, however, the predicted critical pressure associated with global vesicle deformation remains the same as when thermal fluctuations are neglected, approximately six orders of magnitude smaller than the typical collapse pressure recently observed experimentally for GUVs. We conclude by discussing possible sources of this persisting dissonance between theory and experiments.
\end{abstract}

\maketitle

\section{Introduction}

The absolute majority of living cells function optimally in environments with solute concentrations corresponding to an osmotic pressure close to that of physiological saline. Thus, the intracellular environment maintains an osmolarity in the range 250 to 400 mM~\cite{Wennerstrom:PNAS:2020,Oliveberg:QRB:2022}. Nevertheless, in many organisms, the cells are regularly exposed to osmotic stresses induced by differences in the chemical environment between their interior and the surrounding medium. Living organisms have therefore developed a range of mechanisms for coping with these variations~\citep{Bremer:AnnuRevMB:2019,Wood:JGenPhys:2015}. Plants and most bacteria have rigid cell walls that can sustain significant osmotic stresses, and at low external osmotic pressures excessive swelling and membrane rupture are prevented by developing an internal ``turgor pressure''~\citep{Bremer:AnnuRevMB:2019}. For animal cells the situation is more complex, since they have a less rigid plasma membrane associated with their cytoskeleton. Low external osmotic pressures can here lead to membrane rupture, while high external osmotic pressures will cause the cell to shrink in volume resulting in a too crowded internal medium that will slow down metabolic processes~\citep{Harries:2008}. One strategy is to upregulate the intercellular concentration of osmotically active components such as salt or osmolytes, but this comes at a cost in metabolic energy~\citep{Wood:JGenPhys:2015}. A more direct defense against deformation or rupture due to osmotic stress is however built into the intrinsic bending energy of the cellular membrane. In order to  quantify this contribution to vesicle stability against osmotic stress, we will here focus on the osmotic stability of a simple system consisting of unilamellar, spherical vesicles built up by a single-component lipid bilayer. 

\begin{figure}[h]
	\includegraphics[width=110mm,trim={0.5cm 3.5cm 19.0cm 3.0cm},clip]{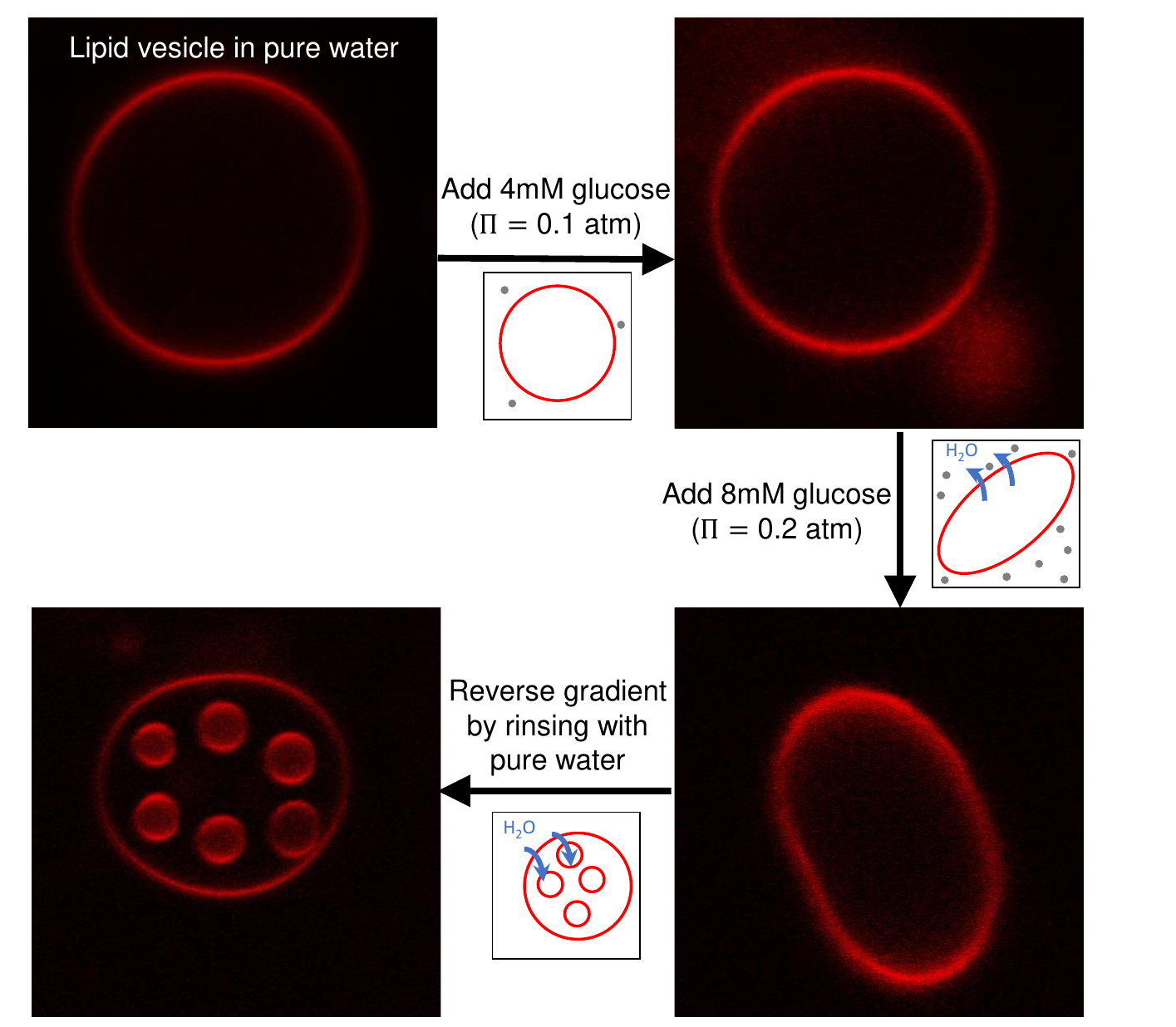} 
	\caption{Microscopy images showing the osmotically induced collapse of a spherical GUV. Glucose is added to the outside solution to induce an osmotic gradient $\Pi$. For small osmotic gradients the vesicles remain effectively spherical, but as the gradient is increased above $\Pi \gtrsim$ 0.15 atm they deform into a prolate shape. Inversion of the gradient by rinsing with pure water after vesicle deformation leads to the formation of well-defined ``daughter vesicles'' through an endocytosis-like process. Images adapted from Ref.~\citep{Sparr:JPCL:2022}.}\label{fig:schematic}
\end{figure}

The osmotic stability of spherical lipid vesicles has been theoretically analysed in the past~\citep{Zhong-can:PRL:1987,Seifert:AdvPhys:1997,Pleiner:PRA:1990}, leading to the conclusion that the contribution from membrane bending to stabilisation against osmotic collapse is relevant only for small (nanometre-scale) lipid vesicles, and negligible for vesicles of sizes comparable to eukaryotic cells ($\sim \SI{10}{\micro\metre}$). In contrast, in a recent study~\citep{Sparr:JPCL:2022} we experimentally measured the osmotically induced deformation of freely suspensded, single-component giant unilamellar vesicles (GUVs) in this size range and found that they were able to sustain significant osmotic stresses before being deformed into a prolate shape at osmotic stresses $\Pi \gtrsim$ 0.15 atm, as shown in Fig.~\ref{fig:schematic}. Upon osmotic reversal, the deformed vesicles exhibited reproducible formation of ``daughter vesicles'' through an endocytosis-like process (fourth step in Fig.~\ref{fig:schematic}) in order to quickly incorporate the external medium, indicating that a substantial bending energy ($>10^3$ times the thermal energy) is stored in the deformed vesicles. These experimental results are broadly in line with previous experimental and computational studies on more complex, multicomponent vesicle systems in buffer solutions, where both osmotic deformation and daughter vesicle formation have been observed~\citep{Pencer:BPJ:2001,Zong:CollSurfB:2018,Dimova:SoftMatter:2020,Saric:SoftMatter:2021,Shibly:BPJ:2016} at roughly similar osmotic gradients as in Ref.~\citep{Sparr:JPCL:2022}. Thus, even though neither of these studies were directly focussed on quantitatively determining the deformation threshold, the overall picture is that GUVs exhibit a significant tolerance towards external osmotic gradients before becoming visibly deformed from spherical shape. Motivated by this discrepancy, we here revisit the problem of the stability of spherical vesicles exposed to external osmotic stresses, taking into account the effect of thermal fluctuations of the lipid membrane. While not previously analysed in the context of osmotic stability, such thermal shape fluctuations have been extensively considered theoretically~\citep{Helfrich:1978,Helfrich:INCD:1984,Helfrich:JPhysFrance:1986,Safran:PRA:1987,Tonchev:PRE:2019,Komura:JPCM:2018} and experimentally~\citep{Hellweg:ACIS:2017,Gradzielski:JCP:2018} and have, among other things, been shown to promote long-ranged repulsive interactions between fluctuating lipid bilayers~\citep{Helfrich:1978,Helfrich:INCD:1984}. Thermal fluctuations have also been theoretically predicted to have a drastic effect on the pressure-induced buckling of nanoscopic solid shells~\citep{Nelson:PNAS:2012}. Our analysis is based on the same basic formalism as used by Ou-Yang and Helfrich in Ref.~\citep{Zhong-can:PRL:1987}, but instead expanding the \emph{thermally} induced deformations in a set of fluctuation modes described by spherical harmonics. Within this framework, we derive a formally exact expression for the configuration integral $Z$ and evaluate it analytically for the two lowest fluctuation modes. Our analysis shows that the inclusion of thermal fluctuations (\emph{i}) shifts the average volume of the vesicle to significantly lower values compared to that of the perfect sphere even for $\Pi = 0$, and (\emph{ii}) changes the vesicle deformation at finite osmotic stresses ($\Pi > 0$) to a continuous one, in contrast to the athermal case. Crucially, however, the collapse of the vesicle volume occurs at the same value of the critical pressure $\Pi_c$ as in the athermal case~\citep{Zhong-can:PRL:1987}. Thus, the inclusion of thermal fluctuations of the vesicle shape within a harmonic approximation cannot explain the significant discrepancies with experimental results, and we conclude the paper by discussing some possible reasons for this unsettling dissonance between theory and experiments. 

\section{Model description}
We consider a unilamellar vesicle formed by a bilayer containing $2N_L$ lipid molecules. The area $a_0$ per molecule is assumed to be constant and independent of the osmotic pressure, thus making the bilayer laterally incompressible and leading to a fixed vesicle area $A_0 = N_L a_0$. For a perfectly spherical vesicle the resulting radius is thus $R_0 = (A_0/4\pi)^{1/2}$. The general expression for the vesicle bending energy $\mathcal{U}_b$ is given to second order by the Helfrich Hamiltonian~\citep{Helfrich:1973}
\begin{equation}\label{eq:Helfrich}
\mathcal{U}_b = \int \left[ \frac{\kappa}{2} (2H-H_0)^2 + \bar{\kappa} K \right] dA,
\end{equation}
where $H$ is the mean curvature, $H_0$ the spontaneous curvature, $\kappa$ the bending rigidity, $K$ the Gaussian curvature, and $\bar{\kappa}$ the saddle-splay rigidity. Notably, compared to previous works, Eq.~\eqref{eq:Helfrich} lacks Lagrange multipliers ensuring that the vesicle area and volume are kept constant. Instead, we handle these conditions in a more mathematically convenient way through a $\delta$ function in the configuration integral, as detailed below. For a vesicle with conserved topology, the Gaussian curvature term is constant and will thus be neglected below. For a symmetric lipid bilayer, such as those composed of only a single lipid component, there is furthermore no spontaneous curvature of the bilayer. In the absence of thermal fluctuations and osmotic stresses, the stable configuration of such a vesicle is that of a perfect sphere with volume $V_0 = 4\pi R^3_0/3 $, with an energy given by $\mathcal{U}_0 = 4\pi (2\kappa + \bar{\kappa})$. At finite temperatures, the vesicle shape will fluctuate around the perfect sphere, leading to an increase in the average membrane bending energy. At constant area, all such thermal excitations must lead to a decrease in vesicle volume, enabled by the fact that the bilayer is permeable to water. This yields an equilibrium vesicle volume $V < V_0$, determined by a balance between the bending energy of the bilayer and the entropy of the thermally excited membrane bending modes.

If the composition of the surrounding medium is changed by the addition or removal of a solute to which the bilayer is impermeable, an osmotic imbalance is created. Assuming ideal solution conditions, the induced osmotic pressure is $\Pi = kT \Delta c$, where $\Delta c$ is the concentration difference, $k$ is Boltzmann's constant, and $T$ the temperature. If the osmotic pressure is higher inside the vesicle than outside it, corresponding to a decrease of solute concentration in the surrounding medium ($\Delta c < 0$), a tension is created in the bilayer, which eventually leads to vesicle rupture. When $\Delta c > 0$, the vesicle instead has a thermodynamic driving force to shrink. If the vesicle interior has a finite solute concentration, osmotic balance will be reached when the vesicle has expelled enough water to equalise the interior and exterior solute concentrations. For the case when the vesicle contains pure solvent, the concentration can however not be equalised by a finite reduction in volume, apart from the small effect of hydronium,  hydroxide and trace amounts of other ions present at micromolar levels even in pure water. One possibility is then that the vesicle collapses completely, thus balancing the external osmotic pressure through a direct monolayer-monolayer interaction in the collapsed state. The second option, which is the one we will consider here, is the balancing of osmotic stress by an increase in bending free energy associated with a finite volume decrease. 

In Ref.~\citep{Zhong-can:PRL:1987}, Ou-Yang and Helfrich analysed the stability of a spherical vesicle subject to an osmotic pressure $\Pi$, using Eq.~\eqref{eq:Helfrich} for the bending energy and neglecting thermal fluctuations. They found that the vesicle remained spherical up to a critical pressure $\Pi_c$, given by
\begin{equation}\label{eq:pc_T0}
    \Pi_c = \frac{2 \kappa}{R_0^3} (6-H_0 R_0).
\end{equation}
For $\Pi > \Pi_c$, an instability occurs in the deformation mode with the largest wavelength, corresponding to a global deformation of the vesicle. A subsequent theoretical study by Pleiner~\citep{Pleiner:PRA:1990} based on a weakly nonlinear stability analysis and including higher-order geometrical correction terms concluded that the deformed state corresponds to an axially symmetric, prolate shape. For a small unilamellar vesicle (SUV) with $R_0 = 50$ nm and bending rigidity $\kappa = 20 kT$, the critical pressure of Eq.~\eqref{eq:pc_T0} with $H_0 = 0$ becomes $\Pi_c \sim 8$ kPa, corresponding to $\Delta c \sim 3$ mM, indicating that vesicles in this size range can withstand non-negligible osmotic stresses without significant deformations. In contrast, a corresponding GUV with $R_0 = \SI{5}{\micro\metre}$ should withstand a pressure of only $8$ mPa, corresponding to $\Delta c \sim 3$ nM, implying that almost any (deliberate or accidental) increase of the solute concentration in the external medium of a GUV suspension would lead to vesicle collapse. As discussed above, this contrasts with the general observation that spherical GUVs can be readily formed and remain stable even in chemically complex environments, and in particular with our recent observation that GUVs remain effectively spherical up to external osmotic pressures $\sim 10^6$ times larger than that predicted by Eq.~\eqref{eq:pc_T0}. 

For small deviations from the sphere, the instantaneous vesicle shape $R(\theta,\phi)$ can be described by an expansion in spherical harmonics $Y_{\ell m}(\theta, \phi)$ with associated coefficients $a_{\ell m}$ ~\citep{Seifert:AdvPhys:1997}:
\begin{equation}\label{eq:R_expansion}
    R(\theta,\phi) = R_0 \left[  1 + \sum_{\ell = 0}^{\infty} \sum_{m=-\ell}^{\ell} a_{\ell m} Y_{\ell m} (\theta,\phi) \right].
\end{equation}
The terms with $\ell = 0$ and $\ell = 1$ correspond, respectively, to a change in average radius and a translation of the vesicle, while the terms with $\ell \geq 2$ describe the shape of the vesicle and thus determine its bending energy. The sum over $\ell$ is truncated above some value $\lmax$ set by the physical constraints of the problem, as further discussed below. To leading order in the coefficients $a_{\ell m}$, the volume and area of the deformed vesicle are given by
\begin{align}
    V &= R_0^3 \left[ \frac{4\pi}{3} \left( 1 + \frac{a_{00}}{\sqrt{4 \pi}} \right)^3 + \sum_{\ell m} a_{\ell m}^2 \right], \label{eq:V_exp} \\
    A &= R_0^2 \left[ 4\pi \left( 1 + \frac{a_{00}}{\sqrt{4 \pi}} \right)^2 + \sum_{\ell m} a_{\ell m}^2 \left(1 + \frac{\ell (\ell+1)}{2} \right) \right], \label{eq:A_exp}
\end{align}
where $\sum_{\ell m}$ implies summation over all $\ell \geq 2$ and the corresponding $m$. The bending energy $\Delta \mathcal{U}_b = \mathcal{U}_b - \mathcal{U}_0$ relative to that of the perfect sphere is furthermore given by~\citep{Seifert:AdvPhys:1997}
\begin{equation}\label{eq:Ubend}
       \Delta \mathcal{U}_b = \frac{\kappa}{2} \sum_{\ell m} a_{\ell m}^2 (\ell + 2) (\ell + 1) \ell (\ell - 1) 
\end{equation}

For small deformations, we have that $a_{00}/\sqrt{4\pi} \ll 1$, and the brackets in Eqs.~\eqref{eq:V_exp}--\eqref{eq:A_exp} can be expanded to leading order in $a_{00}$. Using the constant area constraint, the term in $a_{00}$ in~\eqref{eq:V_exp} can then be eliminated to yield 
\begin{equation}\label{eq:V_simple}
    V \approx R_0^3 \left[ \frac{4\pi}{3} - \frac{1}{4}\sum_{\ell m} a_{\ell m}^2 (\ell + 2) (\ell - 1) \right],
\end{equation}
highlighting that any deformation of the vesicle ($a_{\ell m} \neq 0$) under the constant-area constraint necessarily leads to a \emph{decrease} of the vesicle volume compared to that of the perfect sphere. For later developments, we reexpress~\eqref{eq:V_simple} as an equation for the number of water molecules $N$ in the deformed vesicle:
\begin{equation}\label{eq:N_sphere}
	N \approx N_0 \left[ 1 - \frac{3}{16\pi} \sum_{\ell m} a_{\ell m}^2 (\ell + 2) (\ell - 1) \right], 
\end{equation}
where $N_0$ is the number of water molecules in the perfect sphere. 

\section{Osmotic stability in the absence of thermal fluctuations}
From Eqs.~\eqref{eq:Ubend} and~\eqref{eq:N_sphere}, it follows that the minimum energy for a given $N$ corresponds to a deformation in the longest-wavelength, $\ell = 2$ modes. The bending energy $\Delta \mathcal{U}_b$ relative to the perfect sphere is thus given by 
\begin{equation}\label{eq:Umin}
\Delta \mathcal{U}_b = 16 \pi \kappa \frac{N_0-N}{N_0} \equiv 16 \pi \kappa \hat{N}, 
\end{equation}
where $\hat{N} \equiv (N_0-N)/N_0 \in [0,1]$ quantifies the relative deviation from spherical shape. We now create an osmotic gradient by dissolving solute in the external medium, changing the external chemical potential to $\mu_w < \mu_w^{\theta}$, where $\mu_w^{\theta}$ is the chemical potential of pure water. The (free) energy change $\Delta \mathcal{U}_{\mu}$ relative to the pure water case associated with creating this osmotic imbalance is
\begin{equation}\label{eq:DeltaU_Mu}
\Delta \mathcal{U}_{\mu} = (\mu_w - \mu_w^{\theta})(N_0-N) = -\Pi V_0 \hat{N},
\end{equation}
where we have used the osmotic pressure definition $\Pi V_w = (\mu_w^{\theta} - \mu_w)$, with $V_w$ the molecular volume of water. By combining Eqs.~\eqref{eq:Umin} and~\eqref{eq:DeltaU_Mu}, we note that the total energy change $\Delta \mathcal{U} = \Delta \mathcal{U}_b + \Delta \mathcal{U}_{\mu}$ goes negative when $(\mu_w^{\theta} - \mu_w) > 16\pi \kappa / N_0$, which is equivalent to the stability condition in Eq.~\eqref{eq:pc_T0} with $H_0 = 0$. Thus, due to the linear dependence on $\hat{N}$, for $\Pi > \Pi_c$ the vesicle can always minimise its energy by further decreasing its volume towards zero ($\hat{N} \rightarrow 1$), while for $\Pi < \Pi_c$ the minimum energy always occurs for $\hat{N} = 0$, corresponding to the perfect sphere. In the absence of thermal fluctuations, the vesicle deformation thus corresponds to a sharp transition at $\Pi = \Pi_c$ from a perfect sphere to a collapsed one, as found by Ou-Yang and Helfrich~\citep{Zhong-can:PRL:1987}.

\section{General expression of the configuration integral}
At finite temperature and in the absence of constraints, all the harmonic bending modes in Eq.~\eqref{eq:Ubend} become thermally excited, and each mode contributes an average bending energy of $kT/2$. To avoid an unphysical energy divergence, an upper cutoff $\lmax$ to the expansion in Eq.~\eqref{eq:R_expansion} first needs to be defined. The most straightforward way to estimate $\lmax$ is by assuming that the continuum picture of membrane deformations ceases to be relevant for wavelengths comparable to the bilayer thickness $h$, where $\lmax \sim R_0/h$. For a GUV with $h=5$ nm and $R_0 = $ \SI{5}{\micro\metre}, this estimate yields $\lmax \sim 10^3$ and a total of $\sim 10^6$ excited bending modes, since each value of $\ell$ corresponds to $(2\ell + 1)$ degenerate modes. To account for the relation between volume and bending energy caused by the constraint of constant area $A_0$, we will below treat the volume $V$ or, equivalently, the number of enclosed water molecules $N$ as a dependent variable, while ensuring that the constant-area constraint is obeyed for each individual configuration. This approach is different from previous treatments,~\citep{Seifert:AdvPhys:1997,Safran:PRA:1987} which have instead introduced a free energy term accounting for the area constraint using a virtual (negative) tension in the form of a Lagrange multiplier, ensuring that the constant area condition is satisfied in the mean. 

In the presence of an osmotic gradient, the probability $P_N$ of having a vesicle containing $N$ water molecules is given by
\begin{equation}\label{eq:P_N}
	P_N = \frac{Z_N \exp \left[ \beta (N-N_0)(\mu_w - \mu_w^{\theta}) \right] }{{\displaystyle \sum_{N' = 0}^{N_0} Z_{N'} \exp \left[ \beta (N'-N_0)(\mu_w - \mu_w^{\theta}) \right] }},
\end{equation}
where $\beta = (kT)^{-1}$, $Z_N = e^{-\beta \mathcal{F}_b(N)}$ is the $N$-particle configuration integral and $\mathcal{F}_b(N)$ the corresponding bending free energy, which we define relative to the perfect sphere. Note that $Z_N$, and thus $\mathcal{F}_b(N)$, are independent of the osmotic stress difference, which is fully contained in the second exponential factor of~\eqref{eq:P_N}. 

\begin{figure}[h]
	\includegraphics[width=90mm]{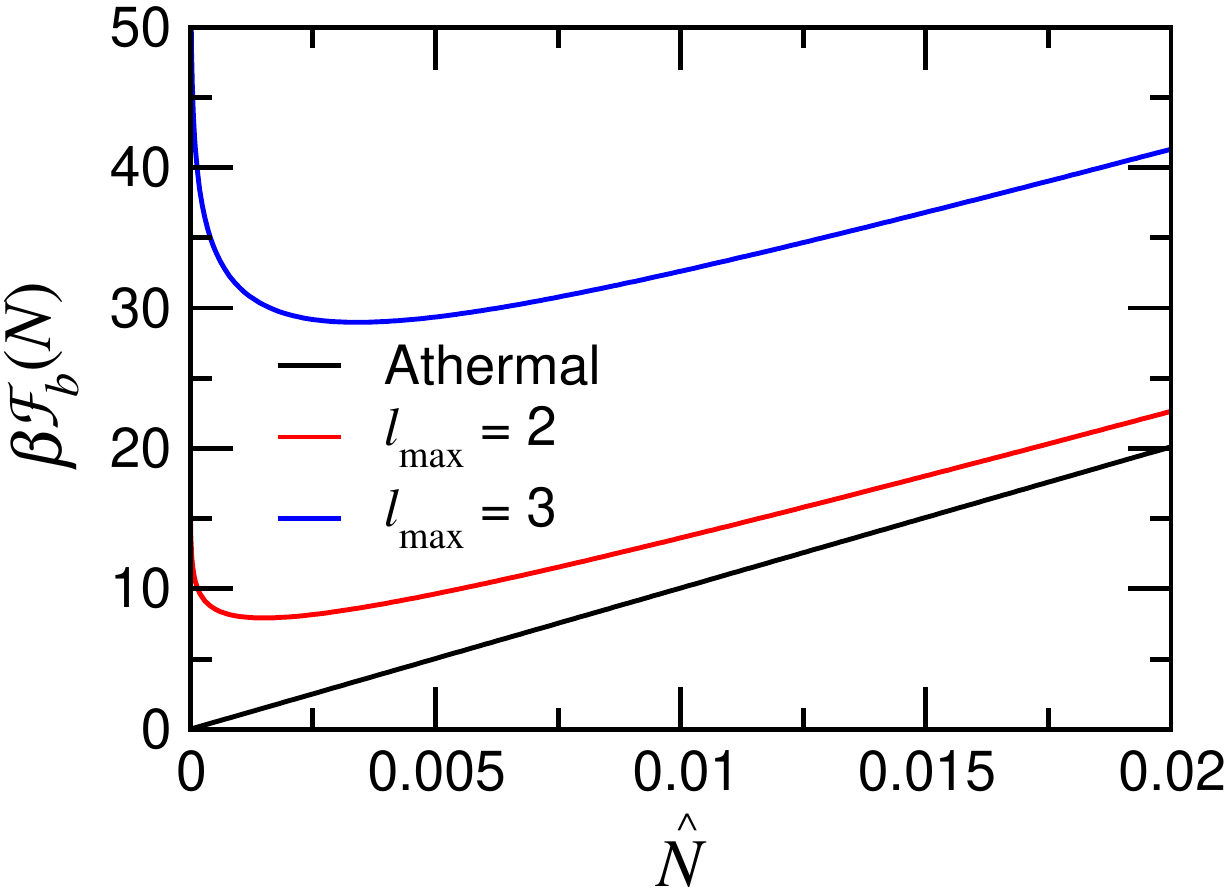} 
	\caption{Bending free energy $\mathcal{F}_b$ in the absence of osmotic stress, plotted as a function of the reduced vesicle volume $\hat{N}$. For the athermal case, the curve shows the linear function in~\eqref{eq:Umin}, which is minimised for $\hat{N} = 0$. For $\lmax = 2$ and $3$, $\mathcal{F}_b(\hat{N})$ is given by the negative logarithms of Eqs.~\eqref{eq:Z2} and~\eqref{eq:Z3}, respectively. Note that, as further fluctuation modes are added, the optimal volume gradually moves away from that of the perfect sphere. }\label{fig:free_energy}
\end{figure}

To find how $\mathcal{F}_b(N)$ varies with vesicle volume, we now proceed to calculate the configuration integral $Z_N$. The fact that $Z_N$ describes fluctuations at constant $N$, corresponding to a constant-volume constraint, introduces a coupling between the different modes through Eq.~\eqref{eq:N_sphere}. We take this condition into account by introducing a $\delta$ function into $Z_N$, yielding
\begin{equation}\label{eq:Z_general}
Z_N = \int \exp [-\beta \Delta \mathcal{U}_b(\{a_{\ell m}\}) ] \delta \left( C_N - \sum_{\ell m} a^2_{\ell m} (\ell+2) (\ell-1) \right) \{ da_{\ell m} \},
\end{equation}
where 
\begin{equation}
C_N \equiv \frac{16 \pi}{3} \hat{N}
\end{equation}
is introduced to simplify notation. We now make the variable substitution $t_{\ell m} \equiv [(\ell + 2)(\ell - 1)]^{1/2}a_{\ell m}$ which allows us to  express $\Delta \mathcal{U}_b$ in Eq.~\eqref{eq:Ubend} as a sum of two parts, where the second one only contains modes with $\ell \geq 3$:
\begin{equation}\label{eq:Ub_simple}
\Delta \mathcal{U}_b = \frac{\kappa}{2} \left[ 6\sum_{\ell = 2}^{\lmax} \sum_{m=-\ell}^{\ell} t^2_{\ell m} + \sum_{\ell = 3}^{\lmax} \sum_{m=-\ell}^{\ell} (\ell^2 + \ell - 6) t^2_{\ell m} \right].
\end{equation}
Since the first sum matches the argument of the $\delta$ function in Eq.~\eqref{eq:Z_general}, the configuration integral simplifies to 
\begin{equation}\label{eq:Z_simple}
\begin{split}
Z_N &= g(\lmax)  e^{-3\hat{\kappa} C_N} \\
&\times \int \exp \left[ -\frac{\hat{\kappa}}{2}\sum_{\ell = 3}^{\lmax} \sum_{m=-\ell}^{\ell} (\ell^2 + \ell - 6) t^2_{\ell m}\right] \delta \left( C_N - \sum_{\ell=2}^{\lmax} \sum_{m=-\ell}^{\ell} t^2_{\ell m} \right) \{ dt_{\ell m} \},
\end{split}
\end{equation}
where $\hat{\kappa} \equiv \kappa/(kT)$, and 
\begin{equation}
g(\lmax) \equiv \prod_{\ell = 2}^{\lmax} [(\ell+2)(\ell-1)]^{-(\ell + \frac{1}{2})} .
\end{equation}
Equation~\eqref{eq:Z_simple} shows explicitly that the lowest allowed energy value is increased by $3 \hat{\kappa} C_N$, yielding a higher reference value for thermal excitations as the vesicle volume is decreased. This is consistent with the analysis of the athermal case, and the value corresponds exactly to $\Delta \mathcal{U}_b$ in Eq.~\eqref{eq:Umin}. To proceed, we note that the integrand in~\eqref{eq:Z_simple} is independent of $m$, and make a variable substitution to $(2\ell+1)$-dimensional polar coordinates, with 
\begin{equation}\label{eq:rho_def}
\rho_{\ell} \equiv \left( \sum_{m=-\ell}^{\ell} t^2_{\ell m} \right)^{1/2}.
\end{equation}
After integrating over the angular coordinates, we get
\begin{equation}\label{eq:Z_final}
\begin{split}
Z_N(\lmax) &= g(\lmax)h(\lmax) e^{-3\hat{\kappa} C_N} \\
& \times \int \exp \left[ -\frac{\hat{\kappa}}{2} \sum_{\ell = 3}^{\lmax} (\ell^2 + \ell - 6) \rho^2_{\ell}\right] \delta \left( C_N - \sum_{\ell=2}^{\lmax} \rho^2_{\ell} \right) \rho_2^4 \cdots \rho_{\lmax}^{2\lmax} d\rho_2 \cdots d\rho_{\lmax},
\end{split}
\end{equation}
with
\begin{equation}\label{eq:h_lmax}
h(\lmax) \equiv \prod_{\ell=2}^{\lmax} \mathcal{A}_{2\ell + 1} = \prod_{\ell=2}^{\lmax} \frac{\pi^{\ell} 2^{\ell+1}}{(2\ell - 1)!!},
\end{equation}
and where $\mathcal{A}_n$ is the surface area of the unit sphere in $n$ dimensions. The $(\lmax-1)$-dimensional integral in Eq.~\eqref{eq:Z_final} represents an exact expression for the configuration integral. In the following, we will proceed to solve it for the cases $\lmax = 2$ and $\lmax = 3$ to yield explicit expressions for the free energy. 

\section{Free energy and vesicle volume in the absence of osmotic stress}
We first consider the case $\lmax = 2$, corresponding to the inclusion of thermal fluctuations only in the most unstable, $\ell = 2$ mode. In this case,~\eqref{eq:Z_final} reduces to a one-dimensional integral that can readily be solved to yield
\begin{equation}\label{eq:Z2}
Z_N(\lmax = 2) = \frac{\pi^2}{24}C_N^{3/2}e^{-3\hat{\kappa}C_N}.
\end{equation}
While also far from realistic values for GUVs, the case $\lmax = 3$ differs qualitatively from $\lmax = 2$, in that it includes the effect on the $\ell = 2$ instability from fluctuations also in the higher, $\ell = 3$ modes. In this case, the configuration integral in~\eqref{eq:Z_final} becomes two-dimensional, and can be straightforwardly solved by subsequent integration over $\rho_2$ and $\rho_3$ to yield
\begin{equation}\label{eq:Z3}
\begin{split}
 Z_N(\lmax = 3) &= \frac{\pi^6 \sqrt{10} }{9.72 \times 10^7}\frac{C_N}{\hat{\kappa}^4} e^{-\frac{9}{2} \hat{\kappa} C_N }
 \\ &\times \left\{ \left[ (3 C_N \hat{\kappa})^2 + 24 C_N \hat{\kappa} \right] I_0\left( \frac{3}{2} C_N \hat{\kappa} \right) - \left[ (3 C_N \hat{\kappa})^2 + 12 C_N \hat{\kappa} + 32 \right] I_1 \left( \frac{3}{2} C_N \hat{\kappa} \right)  \right\},
\end{split}
\end{equation} 
with $I_{\alpha}(x)$ the modified Bessel functions of the first kind and order $\alpha$. By taking the negative logarithm of $Z_N$ in Eqs.~\eqref{eq:Z2} and~\eqref{eq:Z3}, we obtain the corresponding free energies $\mathcal{F}_b(N)$ corresponding to $\lmax = 2$ and $3$, shown in Fig.~\ref{fig:free_energy}. The first obvious effect of increasing $\lmax$ is an increase in the total free energy, as further bending energy modes now become thermally excited. Furthermore, the minimum in free energy is gradually shifted away from the perfect sphere as more fluctuation modes are added. This shift is due to the entropy gain associated with shrinking the vesicle: for $\hat{N} = 0$, corresponding to the perfect sphere, there is only a single possible configuration that simultaneously fulfills the constant-area and constant-volume constraints, leading to a diverging entropy. As $\hat{N}$ increases, the number of accessible vesicle configurations for the given area and volume increases, leading to an entropic driving force for deformation. This is compensated by the bending energy which increases monotonically with $\hat{N}$, together leading to a non-monotonic behaviour of $\mathcal{F}_b(N)$ with a free energy minimum at nonzero $\hat{N}$. 

\begin{figure}[h]
	\includegraphics[width=90mm]{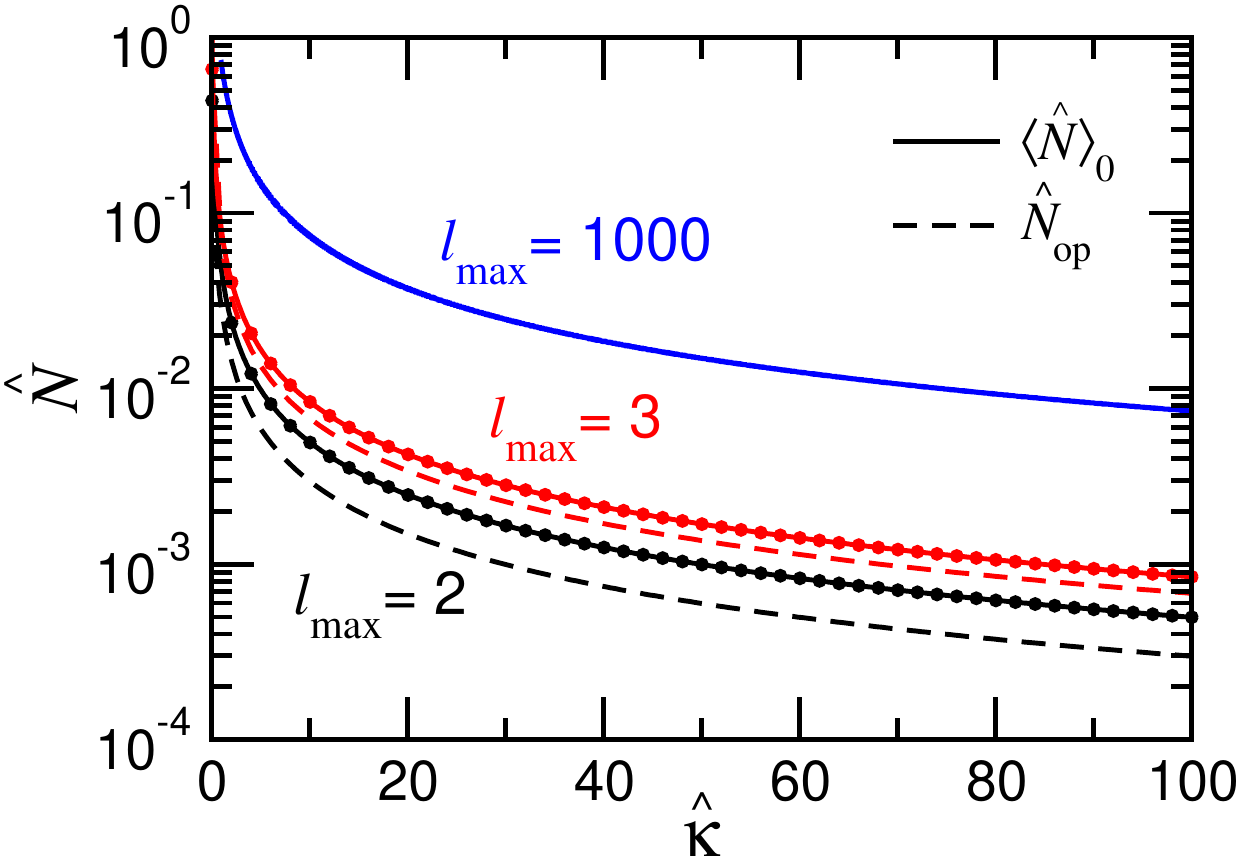} 
	\caption{Average $\langle \hat{N} \rangle_0$ (solid lines/symbols) and optimal $\hat{N}_{\mathrm{op}}$ (dashed lines) reduced vesicle size in the absence of an osmotic gradient, as a function of the reduced bending stiffness $\hat{\kappa}$. Points show results obtained from numerical solution of Eq.~\eqref{eq:N_avg}, and solid lines were obtained from Eq.~\eqref{eq:Navg_EPT}, illustrating the equivalence of the two paths to $\langle \hat{N} \rangle_0$. $\hat{N}_{\mathrm{op}}$ was obtained from numerical minimisation of $\beta \mathcal{F}_b = -\ln Z_N$. }\label{fig:N_kappa}
\end{figure}

From the explicit expressions for $Z_N$ in Eqs.~\eqref{eq:Z2}--\eqref{eq:Z3}, we can quantify the devation from spherical shape in the absence of osmotic stress by calculating the statistical-mechanical average $\langle \hat{N} \rangle_0$ using Eq.~\eqref{eq:P_N} with $\Pi = 0$:
\begin{equation}\label{eq:N_avg}
\langle \hat{N} \rangle_0 = \frac{\displaystyle \int_0^1 \hat{N} Z_{\hat{N}} d\hat{N} }{\displaystyle \int_0^1  Z_{\hat{N}} d\hat{N}},
\end{equation}
where $Z_{\hat{N}}$ is the configuration integral expressed as a function of the reduced particle number $\hat{N}$. 

An alternative, and much simpler, way of obtaining $\langle \hat{N} \rangle_0$ for arbitrary $\lmax$ is to employ the equipartition theorem, which together with Eq.~\eqref{eq:Ubend} yields that the thermal average $\langle a^2_{\ell m} \rangle$ is given by~\citep{Seifert:AdvPhys:1997}
\begin{equation}\label{eq:a2_EPT}
\langle a^2_{\ell m} \rangle = \frac{1}{\hat{\kappa} (\ell+2)(\ell+1)\ell(\ell-1)}.
\end{equation}
Inserting this expression into Eq.~\eqref{eq:N_sphere} and noting that there are $(2\ell + 1)$ degenerate modes for each $\ell$ directly yields an expression for $\langle \hat{N} \rangle_0$:
\begin{equation}\label{eq:Navg_EPT}
\langle \hat{N} \rangle_0 = \frac{3}{16\pi \hat{\kappa}} \sum_{\ell=2}^{\lmax} \frac{2\ell + 1}{\ell(\ell+1)}.
\end{equation}
The monotonically increasing nature of Eq.~\eqref{eq:Navg_EPT} implies that the free energy minimum is gradually shifted towards larger $\hat{N}$ as $\lmax$ is increased. Crucially, $\hat{N}$ diverges logarithmically with $\lmax$ since the summand decays as $\ell^{-1}$ for large $\ell$: this shows that the ($R_0$-dependent) choice of $\lmax$ is crucial when discussing the effect of thermal fluctuations in lipid vesicles. As we show below, using the value $\lmax = 1000$ realistic for GUVs yields physically reasonable values of $\hat{N}$, indicating that this divergence is unproblematic in practice. A comparison between Eq.~\eqref{eq:Navg_EPT} and the numerical solution of Eq.~\eqref{eq:N_avg} shows that the two expressions coincide perfectly for $\lmax = 2$ and 3 for $\hat{\kappa} > 1$, below which the assumption of small deformations breaks down, thus providing an independent check of the expressions~\eqref{eq:Z2}--\eqref{eq:Z3} for the configuration integral. 


Before we proceed to the case of a nonzero osmotic gradient, we note that, for low values of $\lmax$, $\langle \hat{N} \rangle_0$ differs from the \emph{optimal} value $\hat{N}_{\mathrm{op}}$ obtained by instead minimising $\beta \mathcal{F}_b = -\ln Z_N$ with respect to $\hat{N}$, which gives
\begin{equation}\label{eq:N_eq_2}
\hat{N}_{\mathrm{op}} (\lmax = 2) = \frac{3}{32 \pi \hat{\kappa}},
\end{equation}
while, from Eq.~\eqref{eq:Navg_EPT}, we have
\begin{equation}\label{eq:N_avg_2}
\langle \hat{N} \rangle_0 (\lmax = 2) = \frac{5}{32\pi \hat{\kappa}}. 
\end{equation}
The difference between Eqs.~\eqref{eq:N_eq_2} and~\eqref{eq:N_avg_2} seemingly goes against basic statistical-mechanical results, but stems from the fact that we are far from the thermodynamic limit which here corresponds to having a large number of degrees of freedom, \emph{i.e.}, $\lmax \rightarrow \infty$. Physically, this shows that, for very small SUVs where only a few fluctuation modes are thermally excited, fluctuations about the equilibrium volume will be significant. On the other hand, for GUVs with millions of fluctuation modes, $\hat{N}_{\mathrm{op}} \approx \langle \hat{N} \rangle_0$, and thermal fluctuations of the vesicle shape will effectively occur at a constant volume $V_{\mathrm{op}} < V_0$.

In Fig.~\ref{fig:N_kappa}, we present the relative volume reduction $\hat{N}$, using both measures described above, as a function of $\hat{\kappa}$ for $\lmax = 2$ and $3$, together with $\langle \hat{N} \rangle_0$ evaluated from Eq.~\eqref{eq:Navg_EPT} for $\lmax = 1000$, relevant for micron-sized GUVs. For very small values of $\hat{\kappa}$, the volume reduction is clearly too large for the harmonic approximation to be valid, while for physically relevant values of $\hat{\kappa} \sim 20$ and $\lmax = 1000$, we get that $\langle \hat{N} \rangle_0 \sim 4 \times 10^{-2}$, corresponding to a decrease in vesicle volume of $4\%$ compared to that of the perfect sphere. Finally, the results show that the discrepancy between the two volume measures ($\hat{N}_{\mathrm{op}}$ and $\langle \hat{N} \rangle_0$) decreases for $\lmax = 3$ compared to $\lmax = 2$, and will eventually vanish as more fluctuation modes are added. 

\begin{figure}[h]
	\includegraphics[width=90mm]{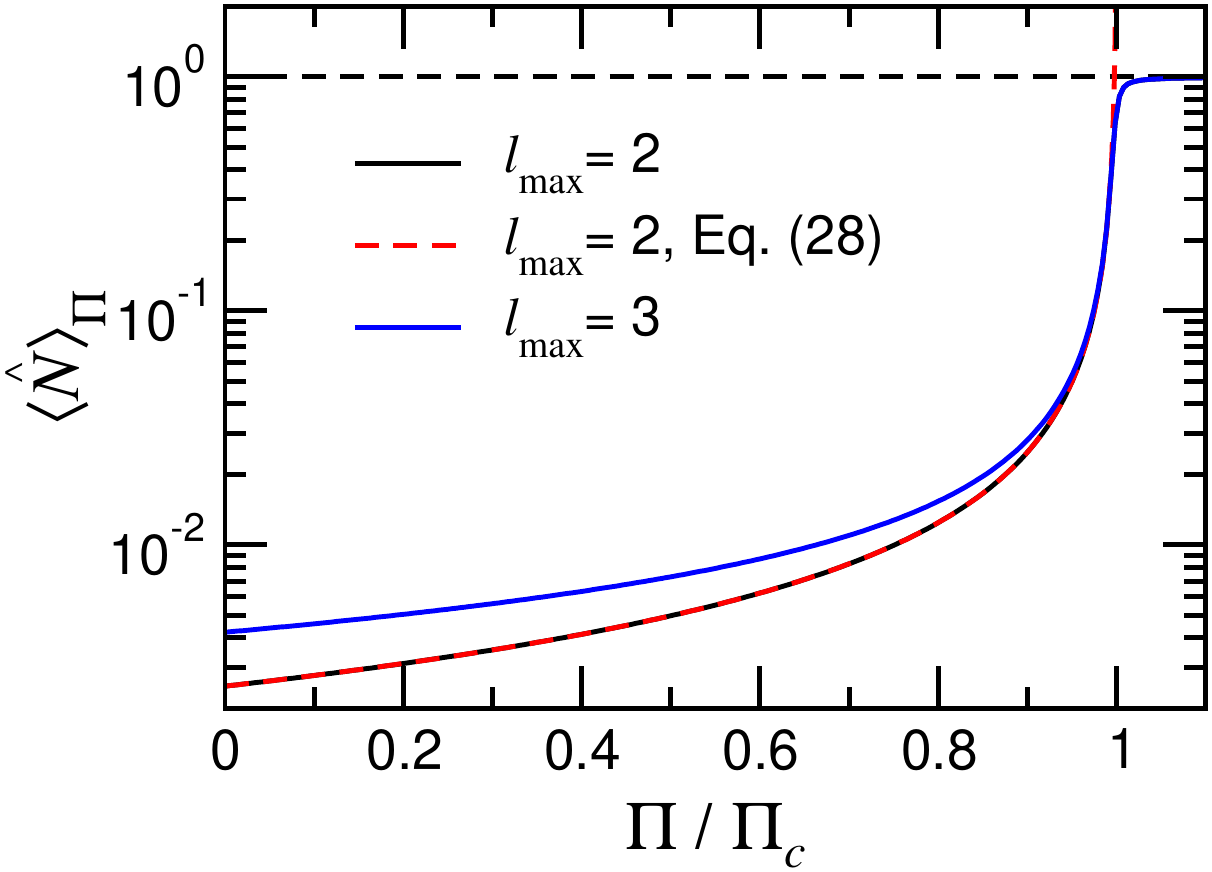}
	\caption{Average reduced vesicle size $\langle \hat{N} \rangle_{\Pi}$ for $\hat{\kappa} = 20$ in the presence of an osmotic gradient $\Pi$, obtained from Eq.~\eqref{eq:N_avg_Pi}. The inclusion of thermal fluctuations makes the vesicle deformation increasingly gradual, while not affecting the location of the vesicle collapse, which is still given by $\Pi_c$ in~\eqref{eq:pc_T0}. The red dashed line shows the simplified expression~\eqref{eq:Navg2}, valid for $\Pi \lesssim \Pi_c$, while the solid lines correspond to an exact evaluation of Eq.~\eqref{eq:N_avg_Pi}. The dashed line for $\hat{N} = 1$ corresponds to the fully collapsed vesicle. }\label{fig:N_Pi}
\end{figure}

\section{Vesicle deformation due to osmotic stress}
We now turn to the case of an imposed osmotic imbalance $\Pi > 0$ across the membrane, corresponding to a higher solute concentration outside than in the vesicle interior. In this case, we can no longer rely on the equipartition theorem to easily express $\langle a^2_{\ell m} \rangle$ as in Eq.~\eqref{eq:a2_EPT}. Instead, we will use the expressions~\eqref{eq:Z2} and~\eqref{eq:Z3} for $Z_N$ to compute the volume reduction in the presence of an osmotic gradient. Reexpressing Eq.~\eqref{eq:P_N} as a function of $\Pi$, we can compute $\langle \hat{N} \rangle_{\Pi}$ according to
\begin{equation}\label{eq:N_avg_Pi}
\langle \hat{N} \rangle_{\Pi} = \frac{\displaystyle \int_0^1 \hat{N} Z_{\hat{N}} \exp(\beta \hat{N} \Pi V_0) d\hat{N}}{\displaystyle \int_0^1  Z_{\hat{N}} \exp(\beta \hat{N} \Pi V_0) d\hat{N}}.
\end{equation}
For $\lmax = 2$, the integrals can be analytically evaluated, leading to the following generalisation of Eq.~\eqref{eq:N_avg_2} valid for $\hat{\kappa} \gtrsim 0.2$ and $\Pi \lesssim \Pi_c$:
\begin{equation}\label{eq:Navg2}
\langle \hat{N} \rangle_{\Pi} (\lmax=2) = \frac{5}{32\pi \hat{\kappa}} \frac{\Pi_c}{\Pi_c-\Pi}, 
\end{equation}
where $\Pi_c$ is still given by the athermal value in Eq.~\eqref{eq:pc_T0}, showing that, for $\lmax = 2$, thermal fluctuations do not affect the \emph{location} of the instability compared to the athermal case, but changes the nature of the transition from abrupt to continuous. The picture does not change qualitatively when we include also the $\ell = 3$ modes into the description, although we here need to numerically evaluate Eq.~\eqref{eq:N_avg_Pi}, leading to the results shown in Fig.~\ref{fig:N_Pi}. Strikingly, the position of the global vesicle deformation remains at $\Pi = \Pi_c$ given by the athermal condition~\eqref{eq:pc_T0} even when fluctuations in $\ell = 3$ modes are included. However, just as in the absence of an osmotic gradient (Fig.~\ref{fig:N_kappa}), the $\hat{N}$ curve is shifted upwards as $\lmax$ is increased from 2 to 3. As $\Pi$ approaches $\Pi_c$ the difference between the two curves decreases, before they collapse in the vicinity of $\Pi_c$. Since the only $\lmax$-dependence in $\langle \hat{N} \rangle_{\Pi}$ (Eq.~\eqref{eq:N_avg_Pi}) comes from the expansion of the free energy $\mathcal{F}_b$, we analyse this effect by forming the difference $\mathcal{F}_b(\lmax) - \mathcal{F}_b(2)$. A direct derivation from Eqs.~\eqref{eq:Z_final} and~\eqref{eq:Z2} gives, after integrating Eq.~\eqref{eq:Z_final} over $\rho_2$:
\begin{equation}\label{eq:F_diff}
\begin{split}
\beta \mathcal{F}_b (\lmax) & - \beta \mathcal{F}_b(\lmax = 2)) \\ 
& \simeq -\ln \int \exp \left[ -\frac{\hat{\kappa}}{2} \sum_{\ell = 3}^{\lmax} (\ell^2 + \ell - 6) \rho^2_{\ell}\right] \left(1- C_N^{-1} \sum_{\ell = 3}^{\lmax} \rho^2_{\ell} \right)^{3/2} \rho_3^6 \cdots \rho_{\lmax}^{2\lmax} d\rho_3 \cdots d\rho_{\lmax}. 
\end{split}
\end{equation}
Equation~\eqref{eq:F_diff} is a monotonically decreasing function of $C_N$ (and thus of $\hat{N}$), showing that the $\ell = 2$ mode becomes increasingly dominant the more the vesicle shape deviates from the perfect sphere. This property of $\mathcal{F}_b$ explains the gradual collapse of the curves for $\lmax = 2$ and 3 as the pressure is increased, and holds for any value of $\lmax$. We thus expect the general features in Fig.~\ref{fig:N_Pi} to hold also for physically realistic values of $\lmax$ for GUVs. In spite of being far from these values, our results furthermore show that the correlation between fluctuations in the most unstable, $\ell = 2$ mode and higher modes does not affect the athermal instability criterion. This observation is far from trivial since when thermally exciting higher bending modes, the energy can be stored in all available modes rather than in only the $\ell = 2$ modes themselves. Such coupled fluctuations however change the nature of the transition from abrupt to continuous: in the absence of an osmotic stress, the average vesicle volume is marginally affected by thermal fluctuations, while for $\Pi \approx \Pi_c$ the effect of thermal fluctuations can be sizeable. Thus, the effect of an osmotic stress is to ``soften'' the bending modes, making them significantly more excited than for $\Pi = 0$. 

\section{Discussion}
By explicitly considering the effect of thermal shape fluctuations, we have shown that spherical vesicles subject to an external osmotic pressure respond in a continuous way by gradually decreasing their volume, in contrast to the abrupt instability occurring in the absence of thermal fluctuations. Our description is formally exact within the harmonic approximation, although the explicit calculations are limited to the small number of fluctuation modes present for $\lmax = 3$. This limitation is due to the difficulty introduced by the constant-volume constraint in Eq.~\eqref{eq:Z_final}, which couples the fluctuations of modes of all $\ell$, and requires further approximations to enable analytical progress towards physically realistic values of $\lmax$. One simple such approximation is to assume that the constant-volume constraint in~\eqref{eq:Z_final} applies only to the $\ell = 2$ mode, thus ignoring the coupling of fluctuations between modes of different $\ell$. This leads to a decoupling of the multidimensional integral into a product of $(\lmax-1)$ Gaussian integrals that can be solved in closed form. This decoupling however means that the position of the minimum in $\mathcal{F}_b(\hat{N})$ is unchanged when modes with $\lmax > 2$ are added, so that we do not capture any additional effects on the vesicle volume compared to the $\lmax = 2$ case. Nevertheless, this approximation could provide useful for obtaining other observables not directly related to the vesicle volume. 

Since our analysis is purely thermodynamic in nature, the results do not depend explicitly on the timescales of equilibration or fluctuations. A rough estimate of the equlibration time can however be obtained from considering the water permeability $\mathcal{P} \approx \SI{16}{\micro \metre \per \second}$ measured for GUVs made of POPC lipids~\citep{Dimova:SoftMatter:2020}. The kinetics of volume change due to a concentration difference $\Delta c$ can however be estimated from the water flux $J_w = \mathcal{P}\Delta c$ and volume change $\dot{V} = J_w A_0 V_w$. The kinetics of the relative volume change $\dot{V}/V_0$ is thus given by
\begin{equation}
\frac{\dot{V}}{V_0} = \frac{3\mathcal{P}\Delta c V_w}{R_0}.
\end{equation}
Using typical values of $\Delta c = 10$ mM and $R_0 = \SI{5}{\micro \metre}$ yields $\dot{V}/V_0 \approx 10\%$ per minute, which indicates that the initial equilibration is a relatively slow process, but nevertheless fully accessible on experimental timescales. Furthermore, once equilibrium has been reached, GUV volume fluctuations are negligible so that the system should be considered ergodic at timescales typical for membrane shape fluctuations at constant volume, which is a significantly faster process. For SUVs, $\dot{V}/V_0$ is 1-2 orders of magnitude larger, and both equilibration and volume fluctuations should occur on timescales of less than seconds. 

The main conclusion of our study is that the vesicle volume formally goes to zero at a well-defined critical pressure $\Pi_c$ identical to the one first derived by Ou-Yang and Helfrich for the athermal case~\citep{Zhong-can:PRL:1987}. As noted above, this critical pressure is many orders of magnitude smaller than the experimentally observed one, indicating that the theoretical description does not include all relevant aspects of the problem. To conclude, we will thus discuss a few possible sources of this significant discrepancy between theory and experiments. 

\begin{enumerate}
\item While exact within the harmonic approximation, one limitation of our treatment is that, as $\Pi$ approaches $\Pi_c$, the vesicle volume gradually decreases beyond the point where the assumption of small deviations from spherical shape implicit in Eqs.~\eqref{eq:R_expansion}--\eqref{eq:N_sphere} becomes invalid. Thus, strictly speaking, the harmonic energy expression of Eq.~\eqref{eq:Ubend} does not cover all of the the experimentally relevant parameter space, suggesting that additional terms should be introduced into the model in order to quantitatively describe the significant deformations at $\Pi \approx \Pi_c$. These terms can come from two different sources: Firstly, Eqs.~\eqref{eq:V_simple} and~\eqref{eq:N_sphere} are valid only to leading order in the amplitudes $a_{\ell m}$. There are thus higher order geometrical corrections to the expressions for the vesicle free energy~\citep{Pleiner:PRA:1990}. Secondly, Eq.~\eqref{eq:Helfrich} is based on a series expansion of the bending energy. Using the planar state as reference, one expects also a fourth-order term with a positive coefficient~\citep{Helfrich:JPhysFrance:1986}. Considering that the bending modes become highly excited under osmotic stress, such terms might give significant contributions to the bending energy. However, the fourth-order bending effects are inversely proportional to $A_0$~\citep{Helfrich:JPhysFrance:1986} and are thus expected to be more important for small vesicles than for GUVs. Interestingly, the inclusion of anharmonic effects in combination with thermal fluctuations have previously been shown to have a drastic effect on the buckling of nanometre-scale solid shells under external pressure~\citep{Nelson:PNAS:2012}, where the critical buckling pressure was however shifted towards \emph{lower} values than in the athermal, harmonic case. 

\item The presence of a concentration difference between the interior and exterior regions of the vesicle creates an asymmetry which gives rise to a nonzero spontaneous curvature $H_0 < 0$ due to solute-bilayer interactions~\citep{Kabalnov:JPC:1995,Lipowsky:ACSNano:2021}, even for the case where the vesicle itself is fully symmetric, leading to an increase of $\Pi_c$ according to Eq.~\eqref{eq:pc_T0}. However, the theoretical $\Pi_c$ corresponds to extremely low concentrations, so that the asymmetry, and thus the spontaneous curvature contribution to the bending energy, should be negligible around $\Pi_c$. However, in the presence of some other stabilising mechanism this spontaneous curvature term might become important for the osmotic pressures where GUVs are experimentally observed to be stable. Unlike the fourth-order bending terms discussed above, according to~\eqref{eq:pc_T0} this effect is expected to be more important for GUVs than for small vesicles, since it scales linearly with $R_0$. 

\item All the results are based on the assumption that the lipid bilayer is laterally incompressible, corresponding to constraining the fluctuations to a constant area $A_0$. While a good approximation, this is not strictly the case and an external osmotic pressure will induce a finite area reduction. The resulting coupling between membrane contraction and osmotic pressure will generically lead to a stabilisation of the spherical vesicle relative to the constant-area case. To leading order, the relative area change $\hat{A} = (A-A_0)/A_0$ should be proportional to the pressure and inversely proportional to the stretching elasticity $K_A$. By dimensional analysis we furthermore expect that $\hat{A} \sim R_0$, implying that the effect should become more significant for GUVs than for small vesicles. 
\end{enumerate}

While the present work provides important insights into the physics of shape fluctuations in vesicles under osmotic stress, significant future work is thus required to analyse to what extent the additional effects discussed above can contribute to reduce or eliminate the discrepancy between theory and experiment, and to further shed light on the role of bending energy as a stabilising mechanism against osmotic stress in biological systems. 

\section{Acknowledgements} 
JS and ES kindly acknowledge funding from the Swedish Research Council (grant IDs 2019-03718 and 2019-05296). 
\bibliography{bibliography}

\end{document}